\documentclass[twocolumn,aps,showpacs]{revtex4-1}
\usepackage[latin9]{inputenc}
\usepackage{amsmath}
\usepackage{graphicx}
\usepackage{esint}
\usepackage{hyperref}
\usepackage{color}
\definecolor{med-blue}{RGB}{25,25,112}
\hypersetup{colorlinks, linkcolor={red},citecolor={blue}, urlcolor={blue}}

\makeatletter
\@ifundefined{textcolor}{}
{%
 \definecolor{BLACK}{gray}{0}
 \definecolor{WHITE}{gray}{1}
 \definecolor{RED}{rgb}{1,0,0}
 \definecolor{GREEN}{rgb}{0,1,0}
 \definecolor{BLUE}{rgb}{0,0,1}
 \definecolor{CYAN}{cmyk}{1,0,0,0}
 \definecolor{MAGENTA}{cmyk}{0,1,0,0}
 \definecolor{YELLOW}{cmyk}{0,0,1,0}
}

\usepackage{color}

\newcommand{\be}{\begin{equation}}
\newcommand{\ee}{  \end{equation}}
\newcommand{\ba}{\begin{eqnarray}}
\newcommand{\ea}{  \end{eqnarray}}
\newcommand{\ket}[1]{\left|#1\right>}

\makeatother

\begin{document}

\title{Level anti-crossings of an NV center in diamond: Decoherence-free
subspaces and 3D sensors of microwave magnetic fields}

\author{K. Rama Koteswara Rao and Dieter Suter}

\affiliation{Fakult{ä}t Physik, Technische Universit{ä}t Dortmund, D-44221
Dortmund, Germany}

\date{\today}
\begin{abstract}
Nitrogen-vacancy (NV) centers in diamond have become an important
tool for quantum technologies. All of these applications rely on long
coherence times of electron and nuclear spins associated with these
centers. Here, we study the energy level anti-crossings of an NV center
in diamond coupled to a first-shell $^{13}$C nuclear spin in a small
static magnetic field. These level anti-crossings occur for specific
orientations of the static magnetic field due to the strong non-secular
components of the Hamiltonian. At these orientations we observe decoherence-free
subspaces, where the electron spin coherence times ($T_{2}^{*}$)
are 5-7 times longer than those at other orientations. Another interesting
property at these level anti-crossings is that individual transition
amplitudes are dominated by a single component of the magnetic dipole
moment. Accordingly, this can be used for vector detection of microwave magnetic
fields with a single NV center. This is particularly important to
precisely control the center using numerical optimal control techniques. 
\end{abstract}


\maketitle

\section{Introduction}

Nitrogen-Vacancy (NV) centers in diamond have many interesting properties
for various applications ranging from quantum information processing
to nano-scale imaging \cite{RevDoherty,RevChildress,RevWalsworth,RevDegen,RevJacques}.
For most of these applications, long coherence times of electron and
nuclear spins associated with the NV center are essential. Dynamical
decoupling pulse sequences are effectively used to decouple NV centers
from their environment and hence improve the coherence times of the
centers \cite{CoryNVDD,HansonNVDD,HansonNVPQC,ShimEPL,JingfuP1q,JingfuP2q,WalsworthNVDD}.
For NV centers, the major source of decoherence is the spin bath formed
by the electron and nuclear spins of impurity atoms (e.g. substitutional
nitrogen) and $^{13}$C nuclear spins in the diamond lattice \cite{MazeDecoh,JW2009PRB,WalsworthDecoh}.
The coherence times of NV centers can be significantly extended in
ultrapure diamond crystals, where the substitutional nitrogen atom
concentration is very low. The nuclear spin bath due to $^{13}$C
can be reduced by using ultrapure diamond crystals enriched with $^{12}$C
atoms \cite{WrachtrupLcoh}. However, $^{13}$C nuclear spins that
are strongly coupled to the electron spin of an NV center can also
be useful as qubits, either as part of a quantum register \cite{Chil2006Sci,Lukin2007Sci,Han2014NatNano,JW2008Sci}
or for storing quantum information \cite{Shim2013}. For example,
the $^{13}$C nuclear spin of the first coordination-shell has a strong
hyperfine coupling with the electron spin of the NV center, which
can be used to implement fast multiqubit gates \cite{Jelez2004,JW2008Sci}.
The disadvantage of using diamond crystals enriched in $^{12}$C is
that these useful qubits are lost.

In this work, we investigate decoherence-free subspaces at energy
level anti-crossings (LACs) of an NV center, whose spin bath is dominated
by the $^{13}$C nuclear spins. LACs of NV centers that occur between
the $m_{s}=0$ and $m_{s}=-1$ spin sublevels of both the ground and
optically excited states have been studied extensively and used for
various purposes \cite{MansonGSLAC,AwschalomLAC,JacquesESLAC,ChilPRA2009},
such as polarizing the nuclear spins. These anti-crossings occur at
magnetic field strengths of $\approx500$ G and $\approx1000$ G.
Here, we study the LACs that occur at much smaller field strengths,
of an NV center coupled to a first-shell $^{13}$C nuclear spin. Specifically,
we study the LACs that occur at two different magnetic field orientations:
(i) The strength and orientation of the magnetic field are such that
the spectral splitting due to the Zeeman interaction of the electron
spin is equal to the splitting due to the hyperfine interaction of
the first-shell $^{13}$C nuclear spin ($\approx127$ MHz). (ii) The
magnetic field is oriented in the plane perpendicular to the N-V axis.
Close to the LACs, the mixing of the states results in ZEro First-Order
Zeeman (ZEFOZ) shift \cite{ZEFOZsellars1,ZEFOZsellars2,ZEFOZsellars3,ZEFOZsuter}
of some of the transitions and correspondingly reduced perturbations
by magnetic-field noise. This effect manifests itself by long coherence
times ($T_{2}^{*}$), almost an order of magnitude longer than at
other orientations. 

An NV center coupled to a first-shell $^{13}$C nuclear spin is particularly
attractive for quantum information processing because of the strong
hyperfine coupling between the electron and nuclear spins \cite{Jelez2004,JW2008Sci,JW2009PRB,Shim2013}.
However, harnessing the full potential of this system requires accurate
knowledge of the Hamiltonian. The time-independent internal Hamiltonian
of this system has been thoroughly investigated \cite{vanWyk1978,Felton2009,Koti2016PRB}.
In addition, precise knowledge of the time-dependent (microwave) MW
Hamiltonian, including the orientation of the MW field with respect
to the center, is also important for precise control of the system.
This information is particularly important in such centers, since
the first-shell $^{13}$C nuclear spin breaks the rotational symmetry
of the center. At the LACs discussed above, the transition amplitudes
of some of the transitions are dominated by a single component of
the magnetic dipole moment. This can be used to determine the strength
and orientation of the MW magnetic field with a single NV center.
A similar vector detection scheme using NV centers was reported in
Ref. \cite{DuVectorMW}. However, that required at least three NV
centers with different orientations in the focal spot of the objective
lens.

This paper is structured as follows. In Sec. \ref{sys}, we discuss
the system, its Hamiltonian and the experimental setup. In Secs. \ref{lac1}
and \ref{lac2}, we analyze the two LACs and discuss the decoherence-free
subspaces and the vector detection of MW magnetic fields. Finally,
in Sec. \ref{conc}, we draw some conclusions.

\section{System and Hamiltonian}

\begin{figure*}[t]
\centering \includegraphics[width=15cm]{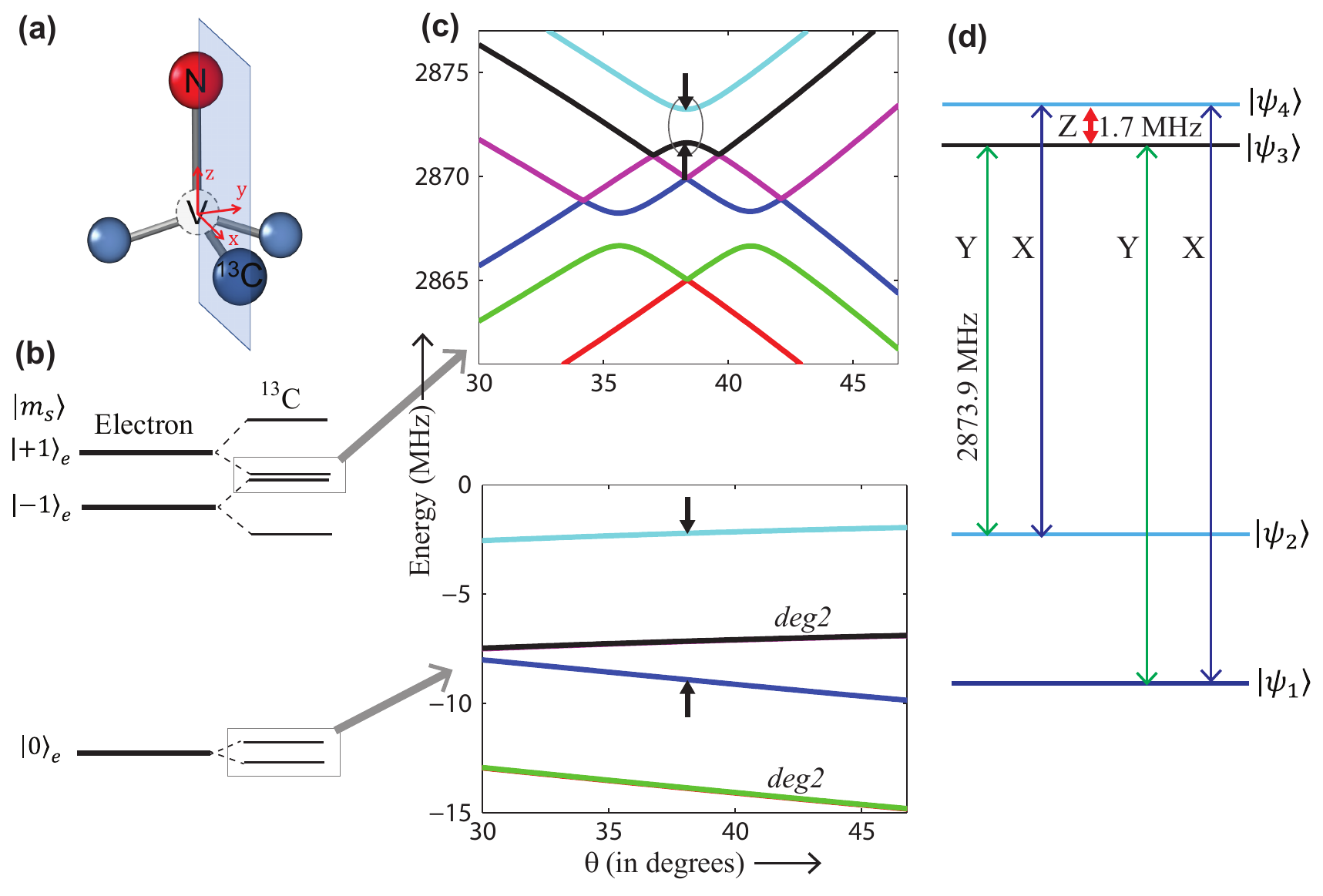} \caption{(a) Structure of an NV center with a $^{13}$C atom in the first coordination
shell. (b) Energy level diagram of the system considering only the
electron and $^{13}$C nuclear spins at the LAC point, $2\gamma_{e}B\cos\theta\approx127$
MHz. (c) Energy levels in gray rectangles of (b) as a function of
$\theta$ of a static magnetic field of strength $B=28.9$ G and $\phi=0^\circ$. Here,
the interaction due to the $^{14}$N nuclear spin is also considered.
The energy levels labeled by \textit{deg2} are doubly degenerate.
(d) Energy levels marked by black arrows in (c) at $\theta=38.4^{\circ}$
(LAC) and the possible electron spin transitions between them.}
\label{Englvl} 
\end{figure*}

\label{sys} The system of interest here is a single NV center coupled
to a first-shell $^{13}$C nuclear spin. Fig. \ref{Englvl}(a) shows
it's structure and defines the coordinate system that we use here.
The NV symmetry axis is the $z$-axis of the center, the $x$-axis
is perpendicular to this axis and lies in the plane containing the
vacancy, nitrogen, and the $^{13}$C atom, and the $y$-axis is perpendicular
to both of them. The Hamiltonian of the total system consisting of
the electron spin ($\mathbf{S}=1$), the $^{13}$C nuclear spin ($\mathbf{I_{1}}=1/2$),
and the $^{14}$N nuclear spin ($\mathbf{I_{2}}=1$) in this coordinate
system can be written as 
\begin{align}
{\cal H}_{\textrm{sys}}= & DS_{z}^{2}+\gamma_{e}\mathbf{B}\cdot\mathbf{S}+\gamma_{n1}\mathbf{B}\cdot\mathbf{I_{1}}+\gamma_{n2}\mathbf{B}\cdot\mathbf{I_{2}}\nonumber \\
 & +PI_{2z}^{2}+\mathbf{S}\cdot\mathcal{A}_{1}\cdot\mathbf{I_{1}}+\mathbf{S}\cdot\mathcal{A}_{2}\cdot\mathbf{I_{2}}.\label{eq:HamSys}
\end{align}
Here, $D=2.87$ GHz is the electron-spin zero-field splitting, and
$\mathbf{B}=B(\sin\theta\cos\phi,\sin\theta\sin\phi,\cos\theta)$
represents the static magnetic field, where $\theta$ and $\phi$
are its polar and azimuthal angles. $P=-4.95$ MHz \cite{Bajaj14Nquadrapole}
represents the quadrupolar splitting of the $^{14}$N nuclear spin,
and $\mathcal{A}_{1}$ and $\mathcal{A}_{2}$ represent hyperfine
tensors of the $^{13}$C and $^{14}$N nuclear spins respectively
with the electron spin. The parameters of the hyperfine tensors are
$\mathcal{A}_{1zz}=128.9$, $\mathcal{A}_{1yy}=128.4$, $\mathcal{A}_{1xx}=189.3$,
and $\mathcal{A}_{1xz}=24.1$ MHz \cite{Koti2016PRB}, and $\mathcal{A}_{2zz}=-2.3$
MHz and $\sqrt{\mathcal{A}_{2xx}^{2}+\mathcal{A}_{2yy}^{2}}=-2.6$
MHz \cite{Mansion14Nhyperfine,Felton2009,PaolaPRB2015}.

The Hamiltonian for the coupling of the MW or radio-frequency (RF)
field to the electron spin transitions can be written as 
\begin{align}
{\cal H}_{\textrm{mw}}= & \sqrt{2}\gamma_{e}B_{\textrm{mw}}(\sin\zeta\cos\eta\ S_{x}+\sin\zeta\sin\eta\ S_{y}\nonumber \\
 & +\cos\zeta\ S_{z})\cos(\omega t+\varphi)),\label{eq:HamMW}
\end{align}
where $B_{\textrm{mw}}$, $\zeta$, and $\eta$ represent the amplitude,
polar, and azimuthal angles respectively of the applied field at the
site of the NV center. $\omega$ and $\varphi$ represent the frequency
and phase of this field. 

All the experiments of this work have been performed using a home-built
confocal microscope for selective excitation and detection of single
NV centers and a MW circuit for resonant excitation of electron spin
transitions. A 20 $\mu$m thin wire was attached to the diamond surface
to generate the MW fields. The used diamond crystal has a natural-abundance
$^{13}$C concentration and the concentration of substitutional nitrogen
centers is $<$ 5 ppb. Studying LACs of the present work requires
a precise orientation of the static magnetic field. This was achieved
by a permanent magnet attached to two rotational stages such that
their axes are orthogonal to each other and cross at the site of diamond
crystal. By rotating the magnet with these rotational stages, a 3D
rotation of the magnetic field can be achieved. The strength of the
magnetic field ($B$) at the site of the NV center was 28.9 G.

\section{Nuclear spin Induced LAC }

\label{lac1}First, consider the LACs that occur in the $m_{s}=\pm1$
manifold when the energy level splitting due to the Zeeman interaction
of the electron spin ($2\gamma_{e}B\cos\theta$) is equal to the corresponding
splitting due to the hyperfine interaction with the $^{13}$C nuclear
spin, which is $\approx127$ MHz. These LACs have been recently used
to study the strong-driving dynamics of a two-level quantum system
beyond the rotating-wave approximation \cite{KotiArXiv2016}. For
the magnetic field of strength 28.9 G, these LACs occur when $\theta$
is close to 38.4$^{\circ}$. The energy level diagram of the system
considering only the electron and $^{13}$C nuclear spins at this
magnetic field orientation is shown in Fig. \ref{Englvl}(b). The
relevant energy levels for the present work are marked with gray rectangles.
Fig. \ref{Englvl}(c) shows these levels on an expanded scale as a
function of $\theta$ of the static magnetic field. As can be seen
from this plot, there are many LACs in the $m_{s}=\pm1$ manifold
when $\theta$ is close to 38.4$^{\circ}$. Here, we analyze only
the LAC at $\theta=38.4^{\circ}$, which is marked by the gray oval.

At this magnetic field orientation, the four energy levels (two in
the $m_{s}=0$ manifold and two in the $m_{s}=\pm1$ manifold) marked
by small black arrows in Fig. \ref{Englvl}(c) are illustrated in
Fig. \ref{Englvl}(d). We label the corresponding eigenstates as $\ket{\psi_{1}}$,
$\ket{\psi_{2}}$, $\ket{\psi_{3}}$, and $\ket{\psi_{4}}$. In the
$\left|m_{s},m_{I_{1}},m_{I_{2}}\right\rangle $basis, they are approximately
\begin{eqnarray}
\left|\psi_{1,2}\right\rangle  & \approx & \left|0,\frac{\left|-\frac{1}{2}\right\rangle \pm\left|\frac{1}{2}\right\rangle }{\sqrt{2}},0\right\rangle ,\nonumber \\
\left|\psi_{3,4}\right\rangle  & \approx & \left|\frac{\left|-1\right\rangle \mp\left|1\right\rangle }{\sqrt{2}},-\frac{1}{2},0\right\rangle .\label{eq:states}
\end{eqnarray}
Between these four energy states, five electron spin transitions are
possible, which are shown by double sided arrows. Four of these five
transitions (thin green and blue arrows) are between the $m_{s}=0$
and $m_{s}=\pm1$ manifolds and they fall in the MW region, and the
fifth transition (thick red arrow), which falls into the RF region,
connects the two states of the $m_{s}=\pm1$ manifold. The four MW
transitions have long coherence times ($T_{2}^{*}$) compared to magnetic
field orientations without LACs. This is because, at the LAC, the
first-order derivatives of these transition frequencies ($\nu_{i}$)
with respect to the magnetic field are zero, i.e., $\frac{\partial\nu_{i}}{\partial B}=\frac{\partial\nu_{i}}{\partial\theta}=\frac{\partial\nu_{i}}{\partial\phi}=0$.
This is known as ZEFOZ shift \cite{ZEFOZsellars1,ZEFOZsellars2,ZEFOZsellars3,ZEFOZsuter}.

Another interesting aspect of these transitions is that they can be
excited only by individual Cartesian components of the MW or RF field.
For the transitions marked by the letter `Y' (green arrows) in Fig.
\ref{Englvl}(d), $\vert\langle\psi_{2}\vert S_{y}\vert\psi_{3}\rangle\vert\approx0.80$
and $\vert\langle\psi_{1}\vert S_{y}\vert\psi_{3}\rangle\vert\approx0.60$
and the corresponding matrix elements of the operators $S_{x}$ and
$S_{z}$ are approximately zero. This implies that these transitions
can be excited only by the $y$-component of the MW field. Note that
from the eigenstates of Eq. \ref{eq:states}, the transition amplitudes,
$\vert\langle\psi_{2}\vert S_{y}\vert\psi_{3}\rangle\vert=\vert\langle\psi_{1}\vert S_{y}\vert\psi_{3}\rangle\vert\approx\frac{1}{\sqrt{2}}$.
The actual difference between these quantities is due to the deviations
from the approximations in Eq. \ref{eq:states}. Similarly, for the
transitions marked by the letter `X' (blue arrows) in Fig. \ref{Englvl}(d),
$\vert\langle\psi_{2}\vert S_{x}\vert\psi_{4}\rangle\vert\approx0.80$
and $\vert\langle\psi_{1}\vert S_{x}\vert\psi_{4}\rangle\vert\approx0.60$
and the corresponding matrix elements of the operators $S_{y}$ and
$S_{z}$ are approximately zero. This implies that these transitions
can be excited only by the $x$-component of the MW field. For the
transition marked by the letter `Z' (Red arrow) in Fig. \ref{Englvl}(d),
$\vert\langle\psi_{3}\vert S_{z}\vert\psi_{4}\rangle\vert\approx1$
and $\vert\langle\psi_{3}\vert S_{x}\vert\psi_{4}\rangle\vert=\vert\langle\psi_{3}\vert S_{y}\vert\psi_{4}\rangle\vert\approx0$,
which implies that this transition can be excited only by the $z$-component
of the RF field. So, in principle, by comparing the experimental transition
amplitudes of these transitions, vector detection of applied RF and
MW fields can be performed.

In the following, we discuss the experiments performed at this LAC.

\begin{figure*}[t]
\centering \includegraphics[width=14cm]{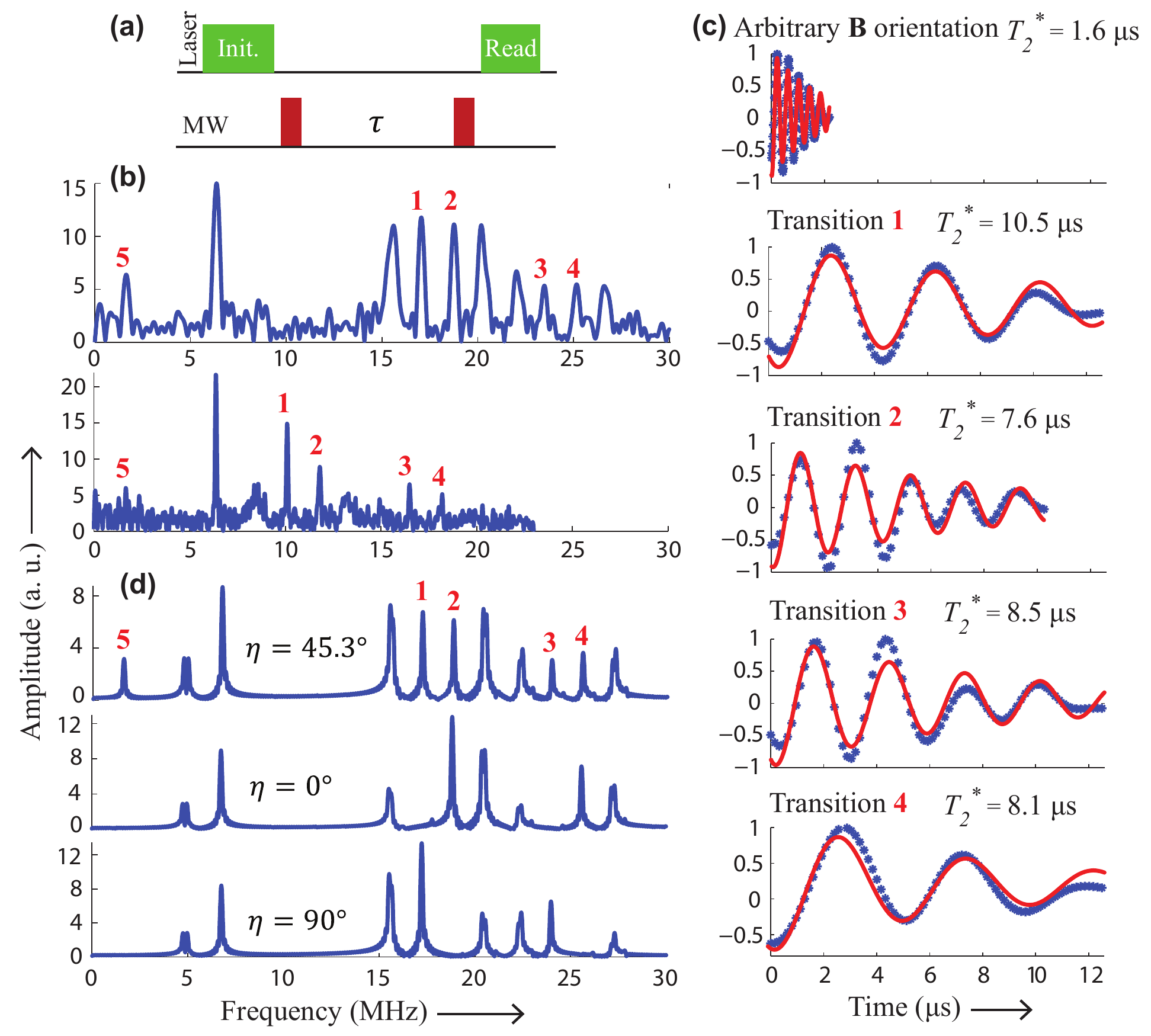} \caption{(a) Pulse sequence to measure FIDs. (b) ESR spectra measured between
the energy levels of Fig. \ref{Englvl}(c). For the upper and lower
spectra, the detuning frequencies ($\nu_{d}$) were 20 and 13 MHz,
and the FID measurement times were 3 and 12 $\mu$s respectively.
(c) FIDs of the transitions 1-4 along with that of an electron spin
transition at an arbitrary $\mathbf{B}$-field orientation. Blue stars
are the data obtained by inverse Fourier transforming the corresponding
spectral lines and the red solid lines fit this data to the equation
$a\cos(2\pi\nu t)\exp(-t/T_{2}^{*})$. (d) Simulated ESR spectra for
different values of $\eta$ of the MW field. The detuning frequency
($\nu_{d}$) for the simulation was 20 MHz.}
\label{Spec} 
\end{figure*}

\textit{Decoherence-free subspaces:} To measure the coherence times
($T_{2}^{*}$) of the transitions discussed above, we recorded the
optically detected Electron Spin Resonance (ESR) spectra at the corresponding
magnetic field orientation ($\theta=38.4^{\circ}$). First, Free Induction
Decays (FIDs) were measured by using the Ramsey sequence shown in
Fig. \ref{Spec}(a), which were then Fourier transformed to get the
frequency domain spectra. The phase of the second MW pulse of the
Ramsey sequence was varied with respect to that of the first one as
$\varphi=-2\pi\nu_{d}\tau$, i.e. as a linear function of the delay
$\tau$ between the pulses. The result is a shift in the measured
spectra by an artificial detuning $\nu_{d}$.

Fig. \ref{Spec}(b) shows the ESR spectra measured between the energy
levels of Fig. \ref{Englvl}(c). The frequency of the applied MW pulses
was 2876.8 MHz. For the upper spectrum, the FID was measured for a
duration of 3 $\mu$s and with a frequency detuning $\nu_{d}$ of
20 MHz. The spectral lines in the frequency range 15-30 MHz correspond
to the single-quantum electron spin transitions between the $m_{s}=0$
and $\pm1$ spin sublevels. Along with these, two more transitions
appear in the spectrum, one at 6.4 MHz, which is a nuclear spin transition
of the $m_{s}=0$ subsystem, and the other at 1.7 MHz, which is an
electron spin transition within the $m_{s}=\pm1$ manifold (marked by thick red arrow in Fig. \ref{Englvl}(d)) \cite{KotiArXiv2016}. These two
transitions appear in the spectrum as zero-quantum transitions \cite{Koti2016PRB},
independent of the detuning $\nu_{d}$. 

For the lower spectrum of Fig. \ref{Spec}(b), the FID was measured
for a duration of 12 $\mu$s and with a frequency detuning of ($\nu_{d}$)
of 13 MHz. The single quantum transitions correspondingly shift by
7 MHz compared to those in the upper spectrum, whereas the zero-quantum
transitions do not. The spectral lines labeled by the numbers 1-5
in both spectra correspond to the five electron spin transitions marked
in Fig. \ref{Englvl}(d). By comparing the two spectra of Fig. \ref{Spec}(b),
it is clear that the transitions labeled by 1-4 have long coherence
times ($T_{2}^{*}$) compared to all the other electron spin transitions.
As discussed earlier, this is due to the ZEFOZ shift at the LAC.

To quantify the $T_{2}^{*}$ of these four transitions, selective
FIDs of them were obtained by inverse Fourier transforming the corresponding
spectral lines. These are shown in Fig. \ref{Spec}(c) in comparison
with that of an electron spin transition at an arbitrary magnetic
field orientation, which doesn't have any LAC. The $T_{2}^{*}$ of
this transition was measured to be 1.6 (1.3, 1.9) $\mu$s and for
the transitions 1-4 the values are 10.5 (8.4, 12.6), 7.6 (5.1, 10.0),
8.5 (6.3, 10.6), and 8.1 (6.5, 9.6) $\mu$s respectively. This corresponds
to an extension of the coherence time by factors of 5-7 .

\textit{Vector detection of the MW field:} For this, we need to determine
the strength ($B_{\mathrm{mw}}$) and orientation (angles $\zeta$
and $\eta$) of the MW field. First, we determine the azimuthal angle
($\eta$) of the MW field, which is the angle between the transverse
component of the MW field and the $x$-axis of the NV center. The
spectral lines labeled by $1$ and $3$ in Fig. \ref{Spec}(b) correspond
to the transitions between the states $\left|\psi_{2}\right\rangle $
and $\left|\psi_{3}\right\rangle $, and $\left|\psi_{1}\right\rangle $
and $\left|\psi_{3}\right\rangle $ respectively. As discussed earlier,
they can be excited only by the $y$-component of the MW field. Similarly,
the spectral lines $2$ and $4$ correspond to the transitions between
the states $\left|\psi_{2}\right\rangle $ and $\left|\psi_{4}\right\rangle $,
and $\left|\psi_{1}\right\rangle $ and $\left|\psi_{4}\right\rangle $,
respectively and they can be excited only by the $x$-component of
the MW field. Therefore, from the amplitudes $I_{\alpha}$ ($\alpha=1,2,3,4$)
of the lines $1\dots4$, the angle, $\eta$ can be determined as follows.

\begin{align}
\vert\tan\eta\vert & =\sqrt{\frac{I_{1}}{I_{2}}}=\sqrt{\frac{I_{3}}{I_{4}}}
\end{align}

For our experimental data, we found $\eta$ $\approx$ 45.3$^{\circ}$.
To test our analysis, we numerically simulated spectra for different
values of $\eta$. Fig. \ref{Spec}(d) shows the resulting spectra.
The top trace, which corresponds to $\eta=45.3^{\circ}$ matches the
experimental spectrum very well. The middle and bottom traces were
simulated for $\eta=0^{\circ}$ and $90^{\circ}$ respectively. In
the middle trace the transitions $1$ and $3$ are absent whereas
in the bottom trace the transitions $2$ and $4$ are absent. Also,
in both of them, the spectral line $5$, which corresponds to the
electron spin RF transition (1.7 MHz) marked by thick red arrow
in Fig.\ref{Englvl}(d), is absent. This is expected, because when
$\eta=0^{\circ}$ or $90^{\circ}$, the MW pulse cannot simultaneously
excite the two transitions ($1$ and $2$ or $3$ and $4$) connecting
the two energy levels of this RF transition with the same $m_{s}=0$
energy level as one of them has zero transition amplitude.

As discussed earlier, the transitions 1 ( $\left|\psi_{2}\right\rangle $
$\leftrightarrow$ $\left|\psi_{3}\right\rangle $) and 5 ( $\left|\psi_{3}\right\rangle $
$\leftrightarrow$ $\left|\psi_{4}\right\rangle $) can be excited
only by the \textit{$y$}- and $z$-components of the MW and RF fields
respectively. This can be used to determine the polar angle ($\zeta$)
of the MW field, which is the angle between the MW field and the $z$-axis
of the NV center. For this, we measured the selective Rabi frequencies
of the transitions 1 and 5, which are 0.44 (for a MW power of 4.72
mW) and 0.36 MHz (for an RF power of 0.67 mW) respectively. The corresponding
expressions can be written as

\begin{eqnarray*}
\sqrt{2}\gamma_{e}B_{mw}\sin\zeta\sin\eta\left|\left\langle \psi_{2}\right|S_{y}\left|\psi_{3}\right\rangle \right| & = & 0.44,\\
\sqrt{2}\gamma_{e}B_{rf}\cos\zeta\left|\left\langle \psi_{3}\right|S_{z}\left|\psi_{4}\right\rangle \right| & = & 0.36\,.
\end{eqnarray*}
Taking the ratios of these two expressions and substituting the values
of $\eta$ and the transition amplitudes ($\vert\langle\psi_{2}\vert S_{y}\vert\psi_{3}\rangle\vert\approx0.80$
and $\vert\langle\psi_{3}\vert S_{z}\vert\psi_{4}\rangle\vert\approx1$),
we get, $\tan\zeta=2.15\frac{B_{rf}}{B_{mw}}$. The angle $\zeta$
was determined by replacing the ratio, $\frac{B_{rf}}{B_{mw}}$ with
the corresponding ratio of square roots of measured MW and RF power
levels. This value is $\zeta=39\textdegree(-3\textdegree,+4\textdegree)$.
The MW and RF power levels were measured before the diamond sample. After
determining the angles, $\zeta$ and $\eta$, it is possible to determine
the amplitudes of the MW and RF fields. From our data, we obtained
them as 0.31 and 0.12 G, respectively. 

\section{Transverse field induced LAC}

\label{lac2} 
\begin{figure}[t]
\centering \includegraphics[width=8.5cm]{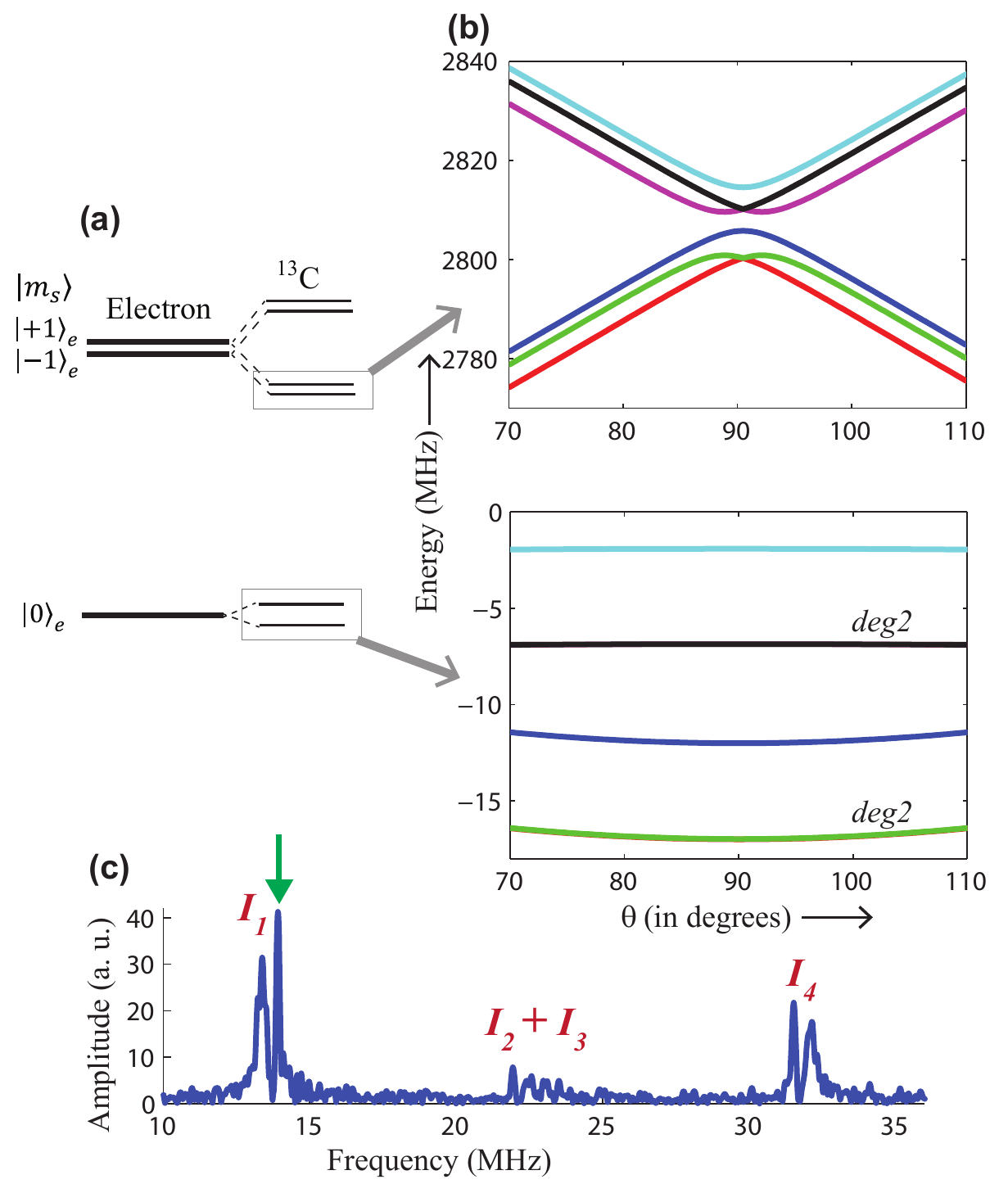} \caption{(a) Energy level diagram of the system considering only the electron
and $^{13}$C nuclear spins when $\theta=90^{\circ}$. (b) Energy
levels in the gray rectangles of (a) as a function of $\theta$ of
the static magnetic field for $\phi=30^{\circ}$and $B=28.9$ G. Here,
the interaction due to the $^{14}$N nuclear spin is also considered.
The energy levels labeled by \textit{deg2} are doubly degenerate. (c)
Experimental ESR spectrum measured between the energy levels of (b)
for $\theta=90^{\circ}$.}
\label{Englvl90} 
\end{figure}

Now, consider the LACs that occur in the $m_{s}=\pm1$
manifold when the static magnetic field is oriented in the transverse
plane of the NV center, i.e., $\theta=90^{\circ}$. The corresponding
energy level diagram of the system considering only the electron and
$^{13}$C nuclear spins is shown in Fig. \ref{Englvl90}(a). The energy
levels marked by gray rectangles are plotted as a function of the
azimuthal angle $\theta$ of the static magnetic field in Fig. \ref{Englvl90}(b),
where the interaction with the $^{14}$N nuclear spin is also included.
From this plot, it is clear that LACs occur in the $m_{s}=\pm1$ manifold
when $\theta=90^{\circ}$. The ESR spectrum measured between these
energy levels for $\theta=90^{\circ}$ and $\phi=30^{\circ}$ is
shown in Fig. \ref{Englvl90}(c).

\textit{Decoherence free subspaces:} Due to the ZEFOZ shift, the spectral
lines have long coherence times ($T_{2}^{*}$) when the $B$-field
is oriented in the $xy$-plane, compared to other orientations. To
compare and quantify the coherence times, we measured the line widths
(full width at half height) of the transition marked by green
arrow in Fig. \ref{Englvl90}(c) as a function of $\theta$ of the
static magnetic field. 
\begin{figure}[h]
\centering \includegraphics[width=8.5cm]{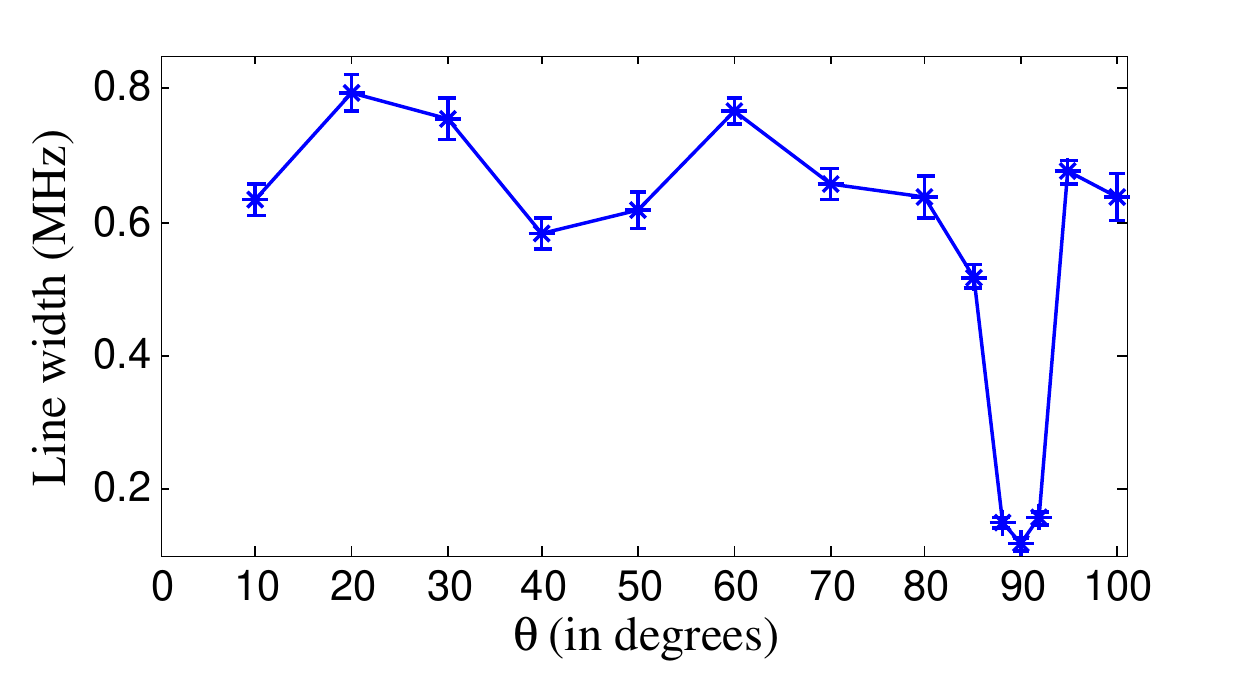} \caption{Line width of the transition marked by green arrow in Fig. \ref{Englvl90}(c)
as a function of $\theta$ of the static magnetic field. Stars connected
by solid line represent the experimental data and the error bars represent
the standard deviation in the measurement of line widths.}
\label{LW} 
\end{figure}

The results are shown in Fig. \ref{LW}. The line widths (of the absolute
value spectra) are in the range 0.60 to 0.80 MHz except when $\theta$
is in between 85 to 95$^{\circ}$, where the line width decreases
sharply and reaches a minimum of 0.12 MHz at $\theta=90^{\circ}$.
This shows that when the static magnetic field is oriented in the
transverse plane, the line width decreases by 5-7 times and hence
the coherence time ($T_{2}^{*}$) increases by the same order. Similar
$T_{2}^{*}$ improvement has been reported in Ref. \cite{WrachtrupESens}
for an NV center without any first-shell $^{13}$C nuclear spin. The
behavior of line width versus $\theta$ of Fig. \ref{LW} is very
similar to the behavior of $1/T_{2}$ versus $\theta$ in Ref. \cite{BajajDecoh}.
There, for similar magnetic field strengths, $T_{2}$ of ensembles
of NV centers was studied as a function of the polar angle $\theta$
of the magnetic field in a diamond sample with a high concentration
($\approx$ 100 ppm) of substitutional nitrogen (p1), where the spin
bath is dominated by the electron spins. An improvement in $T_{2}$
by 2 times was reported when $\theta=90^{\circ}$. In contrast to
these studies, it has been theoretically predicted \cite{MazeDecoh}
and experimentally observed for an ensemble of NV centers \cite{WalsworthDecoh}
that $T_{2}$ is maximum when $\theta=0^{\circ}$ and minimum when
$\theta=90^{\circ}$ in diamond crystals with low concentration of
substitutional nitrogen, where the spin bath is dominated by the $^{13}$C
nuclear spins.

\textit{Determining the azimuthal angle ($\eta$) of the MW field:}
The amplitudes of the spectral lines shown in Fig. \ref{Englvl90}(c)
are labeled as $I_{1}$, $I_{2}$, $I_{3}$, and $I_{4}$. These amplitudes
depend on the azimuthal angle ($\phi$) of the static magnetic field
and also on the angle ($\eta$) between the transverse component of
the MW field and the $x$-axis of the NV center. As discussed in Ref.
\cite{Koti2016PRB}, this dependence can be used to determine the
angle $\eta$. For this purpose, we measured the amplitudes $I_{1}$
to $I_{4}$ as a function of $\phi$ of the static magnetic field
when $\theta=90^{\circ}$. 
\begin{figure}[h]
\centering \includegraphics[width=8cm]{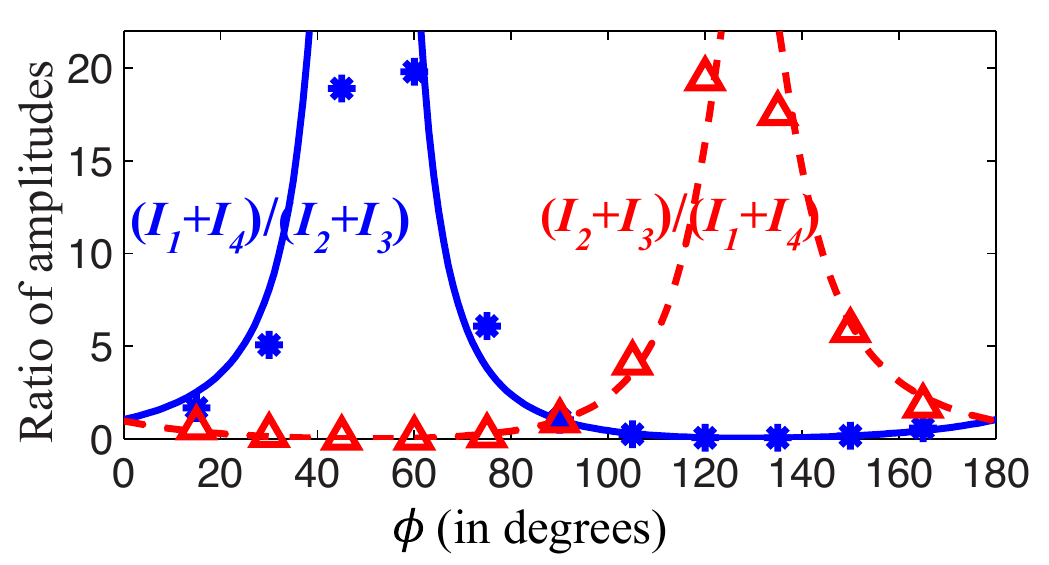} \caption{Ratios of amplitudes of spectral lines as a function of $\phi$ of
the static magnetic field for $\theta=90^{\circ}$.}
\label{sqamps} 
\end{figure}

The experimental data along with the numerically simulated ones are
shown in Fig. \ref{sqamps}. Blue stars and red triangles represent
the experimentally measured quantities $(I_{1}+I_{4})/(I_{2}+I_{3})$
and $(I_{2}+I_{3})/(I_{1}+I_{4})$ respectively. Blue solid and red
dashed lines represent the corresponding simulated quantities. For
the simulation, $\eta=45.3^{\circ}$, which was determined in Sec.
\ref{lac1}, has been used. The simulated quantities are in good agreement
with the experimental ones, which confirms that the angle $\eta$
determined in Sec. \ref{lac1} is correct. 

\section{Conclusion}

\label{conc}

We have studied, experimentally and theoretically, two energy level
anti-crossings in an NV center coupled to a first-shell $^{13}$C
nuclear spin in a small static magnetic field. These anti-crossings
occur in the $m_{s}=\pm1$ manifolds due to the strong non-secular
components of the Hamiltonian for two different static magnetic field
orientations: (i) When the energy level splitting due to the Zeeman
interaction of the electron spin is equal to the splitting due to
the hyperfine interaction of the $^{13}$C nuclear spin ($\approx$
127 MHz). (ii) When the magnetic field is oriented in the transverse
plane of the NV center. At both of these LACs, we observed decoherence
free subspaces due to the ZEFOZ shift, i.e., the coherence times ($T_{2}^{*}$)
of some of the transitions are up to 7 times longer than those at
other orientations of the magnetic field. At these LACs, some of the
electron spin transition amplitudes are dominated by a single component
of the magnetic dipole moment. This has been used to perform vector
detection of the MW magnetic field by a single NV center. The azimuthal
angle of the MW field with respect to the atomic structure of the
center has been determined accurately, but the accuracy of the determined
polar angle of the MW field is limited. This is due to the impedance
mismatches in the MW circuit caused by the copper wire attached to
the diamond surface by which the MW fields are generated. However,
with a near perfect impedance matching MW circuit, the polar angle
can be determined more accurately. The presented scheme will be useful for vector
microwave magnetometry by a single spin. Determining the orientation of
the MW field is also important to precisely control the NV center using
optimal control techniques as the center is not symmetric with respect
to the NV axis due to the presence of \textsuperscript{13}C atom
in the first-shell.
\begin{acknowledgments}
We thank Jingfu Zhang for useful discussions. This work was supported
by the DFG through grant no. Su 192/31-1.
\end{acknowledgments}

\bibliography{bibNV1}

\end{document}